\documentclass[journal,twocolumn]{IEEEtran}
\usepackage{amsfonts}
\usepackage{amsmath}
\usepackage{amssymb}
\usepackage{algorithm}
\usepackage{algorithmic}
\usepackage{bm}
\usepackage{cite}
\usepackage{color}
\usepackage{dsfont}
\usepackage{enumerate}
\usepackage{epsfig}
\usepackage{epstopdf}
\usepackage{graphicx}
\usepackage{mathrsfs}
\usepackage{mathtools}
\usepackage{relsize}
\usepackage[T1]{fontenc}
\usepackage{caption}
\usepackage{etoolbox}
\def\BD{\begin{displaymath}}
\def\BE{\begin{equation}}
\def\BEA{\begin{eqnarray}}
\def\BEAs{\begin{eqnarray*}}
\def\ED{\end{displaymath}}
\def\EE{\end{equation}}
\def\EEA{\end{eqnarray}}
\def\EEAs{\end{eqnarray*}}

\def\bA{{\bf A}}
\def\ba{{\bf a}}
\def\bB{{\bf B}}

\def\bC{{\bf C}}

\def\bd{{\bf d}}
\def\bE{{\bf E}}
\def\be{{\bf e}}
\def\bF{{\bf F}}
\def\bG{{\bf G}}

\def\bH{{\bf H}}

\def\bI{{\bf I}}

\def\bn{{\bf n}}
\def\bP{{\bf P}}

\def\bR{{\bf R}}

\def\bU{{\bf U}}

\def\bV{{\bf V}}

\def\bW{{\bf W}}

\def\bX{{\bf X}}
\def\bx{{\bf x}}

\def\by{{\bf y}}
\def\bZ{{\bf Z}}

\def\b_eta{\mbox{\boldmath $\eta$}}

\def\bnu{\mbox{\boldmath $\nu$}}

\def\diag { {\rm diag} }

\def\trace{ {\rm tr} }

\def\L2{L^{2}[-{\pi \over 2} , {\pi \over 2}]}

\def\bnu{\bnu}

\DeclarePairedDelimiterX\set[1]\lbrace\rbrace{#1}
\graphicspath{{./}{Figures/}}

\newtheorem{proposition}{Proposition}

\begin{document}
\title{Channel-Statistics-Based Hybrid Precoding for Millimeter-Wave MIMO Systems With Dynamic Subarrays}

\author{Juening~Jin,
~Chengshan~Xiao,~\IEEEmembership{Fellow,~IEEE,}
~Wen~Chen,~\IEEEmembership{Senior Member,~IEEE,}
~Yongpeng~Wu,~\IEEEmembership{Senior Member,~IEEE}
\vspace{-1cm}
\thanks{The work of C. Xiao was supported in part by US National Science Foundation under Grants ECCS-1827592. The work of W. Chen was supported in part by National Natural Science Foundation of China under Grant 61671294, STCSM Key Fundamental Project under Grant 16JC1402900 and 17510740700, National Science and Technology Major Project under Grant 2018ZX03001009-002. The work of Y. Wu is supported in part by  the National Science Foundation (NSFC) under Grant 61701301
and Young Elite Scientist Sponsorship Program by CAST. Part of this work has been carried out while J. Jin was visiting Missouri University of Science and Technology. Part of the material in this paper was presented at the IEEE ICC, Kansas City, MO, USA, 2018. (\textit{Corresponding author: Yongpeng Wu}.)}
\thanks{J. Jin and W. Chen are with Shanghai Institute of Advanced Communications and Data Sciences, the Department of Electronic Engineering, Shanghai Jiao Tong University, Shanghai 200240, China (E-mail:
jueningjin@gmail.com, wenchen@sjtu.edu.cn).}
\thanks{C. Xiao is with the Department of Electrical and Computer Engineering,
  Lehigh University, Bethlehem, PA 18015, USA
 (E-mail: xiaoc@lehigh.edu).}
\thanks{Y. Wu is with the Department of Electronic Engineering, Shanghai Jiao Tong University, Shanghai 200240, China (E-mail: yongpeng.wu@sjtu.edu.cn).}
}
\maketitle
\vspace{-1cm}
\begin{abstract}
This paper investigates the hybrid precoding design for millimeter wave (mmWave) multiple-input multiple-output (MIMO) systems with finite-alphabet inputs. The mmWave MIMO system employs partially-connected hybrid precoding architecture with dynamic subarrays, where each radio frequency (RF) chain is connected to a dynamic subset of antennas. We consider the design of analog and digital precoders utilizing statistical and/or mixed channel state information (CSI), which involve solving an extremely difficult problem in theory: First, designing the optimal partition of antennas over RF chains is a combinatorial optimization problem, whose optimal solution requires an exhaustive search over all antenna partitioning solutions; Second, the average mutual information under mmWave MIMO channels lacks closed-form expression and involves prohibitive computational burden; Third, the hybrid precoding problem with given partition of antennas is nonconvex with respect to the analog and digital precoders. To address these issues, this study first presents a simple criterion and the corresponding low complexity algorithm to design the optimal partition of antennas using statistical CSI. Then it derives the lower bound and its approximation for the
average mutual information, in which the computational complexity is greatly reduced compared to
calculating the average mutual information directly. In addition, it also shows that the lower bound with a constant shift offers a very accurate approximation to the average mutual information. This paper further proposes utilizing
the lower bound approximation as a low-complexity and accurate alternative for developing a manifold-based gradient ascent algorithm to find near optimal analog and digital precoders. Several numerical results are provided to show that our proposed algorithm outperforms existing hybrid precoding algorithms.
\end{abstract}
\vspace{-0.2cm}
\begin{IEEEkeywords}
Hybrid precoding, finite-alphabet inputs, matrix factorization, nonconvex optimization.
\end{IEEEkeywords}
\vspace{-0.2cm}
\section{Introduction}
Massive multiple-input multiple-output (MIMO) systems operating in the Millimeter wave (mmWave) band is a key technique candidate for future generation cellular systems to address the wireless spectrum crunch. It makes use of the frequency band from 30 GHz to 300 GHz, which provides a much wider bandwidth than current cellular systems operating in microwave bands. In addition, a short wavelength of radio signals in the mmWave band enables very large antenna arrays to be equipped at the transceivers, and this can provide significant increase of the spectral efficiency.

For mmWave MIMO systems, hybrid analog and digital precoding architectures have been proposed to achieve high spectral efficiency with low cost and power consumption. Extensive work has been devoted to designing hybrid precoding algorithms under perfect channel state information (CSI) and different constraints \cite{el2014spatially,zhang2014achieving,yu2016alternating,sohrabi2016hybrid,rusu2016low,rajashekar2016hybrid,lopez2016hybrid,rajashekar2017iterative}. However, it is difficult to obtain the perfect CSI in mmWave MIMO systems. The reason is that the channel matrix measured at the baseband cannot be obtained directly because it is intertwined with the choice of analog precoders. Furthermore, conventional MIMO channel estimation is incapable of utilizing array gain in mmWave systems, and it leads to low signal-to-noise ratio (SNR). Therefore, the conventional channel estimation requires long training sequences to estimate mmWave MIMO channels, which is impractical due to fast variation of mmWave MIMO channels.

To address the challenge of training overhead, \cite{alkhateeb2013hybrid} proposed a hybrid precoding algorithm for single-user MIMO systems with partial knowledge of the CSI. For the multi-user MIMO scenario, \cite{li2017optimizing} devised a mix-CSI-based hybrid precoding structure, where the analog precoding design is based on the slow-varying channel statistics, and the digital precoding design is based on the instantaneous CSI. Then the dimension of the effective channel matrix (instantaneous CSI) is greatly reduced. However, the work in \cite{alkhateeb2013hybrid} and \cite{li2017optimizing}  considered only the fully-connected
hybrid architecture, which requires much more phase shifters compared to the partially-connected structure \cite{han2015large}. In the partially-connected structure, the antenna array is partitioned into a number of smaller disjoint subarrays, each of which is driven by a single radio frequency (RF) chain\cite{karamalis2006adaptive}. This structure is an extension of classic antenna selection methods, which allocate each RF chain to an antenna element \cite{sanayei2004antenna}. In \cite{gao2016energy}, the authors developed a successive interference cancellation based hybrid precoding for partially-connected structure with fixed subset of antennas. The partially-connected structure with dynamic subset of antennas is considered in \cite{park2017dynamic}, and a low complexity greedy algorithm is also proposed to design the best partitioning/grouping of antennas over the RF chains.

Furthermore, most existing works on hybrid precoding assume Gaussian inputs, which are rarely realized in practice. It is well known that practical systems utilize finite-alphabet inputs, such as phase-shift keying (PSK) or quadrature amplitude modulation (QAM). Precoding designs under Gaussian inputs have been shown to be quite suboptimal for practical systems with finite-alphabet inputs \cite{guo2005mutual,lozano2006optimum,xiao2008mutual,perez2010mimo,xiao2011globally,zeng2012low,wu2015linear,wu2017low,wu2017secure,jin2017linear,jin2017generalized,wu2018survey}. Recently, the authors in \cite{jin2018hybrid} presented a Broyden-Fletcher-Goldfarb-Shanno based hybrid precoding algorithm for mmWave MIMO systems with finite-alphabet inputs. The proposed algorithm utilizes both gradient and Hessian information, and simulation results showed that it outperforms existing hybrid precoding algorithms including \cite{yu2016alternating,rusu2016low,rajashekar2016hybrid,rajashekar2017iterative}.

\subsection{Contributions}
In this paper, we investigate the hybrid precoding design for mmWave MIMO systems with finite-alphabet inputs under the following assumptions: 1) the system employs partially-connected hybrid precoding structure with dynamic subset of antennas; 2) the partition of antennas and analog precoder are designed based on statistical CSI, and the digital precoder is designed based on either statistical CSI or instantaneous CSI. We consider the statistical-CSI-based scenario and the mixed-CSI-based scenario, and the corresponding hybrid precoding problems under two scenarios have the same mathematical form. Then we propose a manifold-based gradient ascent algorithm to solve the hybrid precoding problem. The contributions of this paper are summarized as follows:
\begin{itemize}
  \item We present a simple criterion to design the best partition of antennas using statistical CSI. The corresponding dynamic subarray design is a (nonconvex) combinatorial optimization problem, and we propose a low complexity algorithm to solve this problem.
  \item We derive a lower bound of the average mutual information for mmWave MIMO channels. The lower bound plus a constant shift serves as a very accurate approximation to the average mutual information, and its complexity is much lower than the original average mutual information. To further reduce the complexity, we also derive an accurate approximation of the proposed lower bound.
  \item We propose a manifold-based gradient ascent algorithm to design hybrid precoders. Simulation results show that 1) the proposed algorithm converges to a near globally optimal solution from arbitrary initial points; 2) the performance of mixed-CSI-based hybrid precoding is very close to that of instantaneous-CSI-based hybrid precoding. 3) the statistical-CSI-based hybrid precoding can achieve higher energy efficiency than the fully-connected hybird precoding.
\end{itemize}

\subsection{Notations}
The following notations are adopted throughout the paper: Boldface lowercase letters, boldface uppercase letters, and calligraphic letters are used to denote vectors, matrices and sets, respectively. The real and complex number fields are denoted by $\mathds{R}$ and $\mathds{C}$, respectively. The superscripts $(\cdot)^{\mathrm{T}}$, $(\cdot)^{*}$ and $(\cdot)^\mathrm{H}$ stand for transpose, conjugate, and conjugate transpose operations, respectively. $\trace(\cdot)$ is the trace of a matrix; $\|\!\cdot\!\|$ denotes the Euclidean norm of a vector; $\|\!\cdot\!\|_F$ represents the Frobenius norm of a matrix; $E_{\bx}(\cdot)$ represents the statistical expectation with respect to $\bx$; $\bX_{kl}$ represents the $(k,l)$-th element of $\bX$; $\bI$ and $\bm{0}$ denote an identity matrix and a zero matrix, respectively, with appropriate dimensions; $\circ$ represents the Hadamard matrix product; $\mathcal{I}(\cdot)$ represents the mutual information; $\Re$ and $\Im$ are the real and imaginary parts of a complex value; $\log(\cdot)$ is used for the base two logarithm.

\section{System and Channel Models}
In this section, we present system and channel models for mmWave MIMO systems.
\subsection{System Model}
Consider a point-to-point mmWave MIMO system, where a transmitter with $N_\mathrm{t}$ antennas sends $N_\mathrm{s}$ data streams to a receiver with $N_\mathrm{r}$ antennas. The number of RF chains at the transmitter is $N_\mathrm{rf}$, which satisfies $N_\mathrm{s}\!\leq\! N_\mathrm{rf}\!\leq\! N_\mathrm{t}$. We consider the hybrid precoding scheme, where $N_\mathrm{s}$ data streams are first precoded using a digital precoder, and then shaped by an analog precoder. The received baseband signal $\by\!\in\! \mathds{C}^{N_\mathrm{r}\times 1}$ can be written as
\begin{align}\label{I1}
\by=\bH\bF\bB\bx\!+\!\bn
\end{align}
where $\bH\!\in\!\mathds{C}^{N_\mathrm{r}\times N_\mathrm{t}}$ is the mmWave channel matrix; $\bF\!\in\! \mathds{C}^{N_\mathrm{t}\times N_\mathrm{\mathrm{rf}}}$ is the analog precoder; $\bB\!\in\! \mathds{C}^{N_\mathrm{rf}\times N_\mathrm{s}}$ is the digital precoder; $\bx\!\in\! \mathds{C}^{N_\mathrm{s}\times 1}$ is the input data vector and $\bn\!\in\!\mathds{C}^{N_\mathrm{r}\times 1}$ is the independent and identically distributed (i.i.d.) complex Gaussian noise with zero-mean and covariance $\sigma^2\bI$. To simplify our system model, we omit the analog and digital combiner, which can be designed similarly as the hybrid precoder.

In this paper, the analog precoder $\bF$ is implemented by a dynamic phase shifter subarray, where each RF chain is connected to a dynamic subset of transmit antennas. Let $\mathcal{S}_j$ denote the collection of transmit antennas connected to $j$th RF chain. We partition $N_\mathrm{t}$ transmit antennas into $N_\mathrm{rf}$ subsets satisfying
\begin{align}\label{I2}
\mathcal{S}\!=\!\left\{\{\mathcal{S}_j\}_{j=1}^{N_\mathrm{rf}}\left|
\begin{aligned}
\bigcup_{j=1}^{N_\mathrm{rf}}\mathcal{S}_j\!=\!\left\{1,2,...,N_\mathrm{t}\right\},\mathcal{S}_j \cap \mathcal{S}_k\!=\!\emptyset,\;\;\forall j\!\neq\! k
\end{aligned}
\right.\right\}.
\end{align}
Since each RF chain can be connected to different number of antennas, the cardinalities of $\{\mathcal{S}_j\}_{j=1}^{N_\mathrm{rf}}$ are different. In addition, if the $i$th transmit antenna is connected to the $j$th RF chain, i.e.,$i\!\in\!\mathcal{S}_j$, the $(i,j)$th entry of $\bF$ has unit modulus, otherwise it is zero. Therefore, the constraints on $\bF$ can be expressed by
\begin{align}\label{I3}
|\bF_{\!ij}|\!=\!\mathbf{1}_{\mathcal{S}_j}(i),\;\;\forall (i,j)
\end{align}
where $\mathbf{1}_{\mathcal{S}_j}(i)$ is the indicator function:
\begin{align}
\mathbf{1}_{\mathcal{S}_j}(i)\!=\!\left\{
\begin{aligned}
&1\quad\quad \mathrm{if}\;\;i\!\in\!\mathcal{S}_j \\
&0\quad\quad \mathrm{otherwise}.
\end{aligned}
\right.
\end{align}

The transmitted signal is restricted by a total power constraint $P$:
\begin{align}\label{I4}
E_{\bx}\left\|\bF\bB\bx\right\|^2\!=\!\trace\left(\bB^H\bF^H\bF\bB\right)\!\leq\!P.
\end{align}
To decouple $\bF$ and $\bB$ in coupled power constraint \eqref{I4}, we consider the following change of variables:
\begin{align}
&\bar{\bF}\!=\!\bF(\bF^H\bF)^{-\frac{1}{2}}\\
&\bar{\bB}\!=\!(\bF^H\bF)^{\frac{1}{2}}\bB.
\end{align}
Then the power constraint in \eqref{I4} becomes
\begin{align}\label{I5}
\mathcal{B}\!=\!\left\{\bar{\bB}\left|\trace\left(\bar{\bB}^H\bar{\bB}\right)\!\leq\!P\right.\right\}
\end{align}
and the constraints on $\bar{\bF}$ can be expressed by
\begin{align}
\mathcal{F}\!=\!\left\{\bar{\bF}\left|\left|\bar{\bF}_{\!ij}\right|\!=\!|\mathcal{S}_j|^{-\frac{1}{2}}\mathbf{1}_{\mathcal{S}_j}(i),\;\forall (i,j)\right.\right\}.
\end{align}
Furthermore, by plugging $\bar{\bF}$ and $\bar{\bB}$ into the system model in \eqref{I1}, we have
\begin{align}\label{I6}
\by=\bH\bar{\bF}\bar{\bB}\bx\!+\!\bn.
\end{align}
Combining \eqref{I5} and \eqref{I6}, we observe that $\bH\bar{\bF}$ and $\bar{\bB}$ can be regarded as the effective channel and precoder for typical MIMO Gaussian channels, respectively. Since there exists a one-to-one mapping between $(\bF,\bB)$ and $(\bar{\bF},\bar{\bB})$, we will focus on designing the effective analog and digital precoders $(\bar{\bF},\bar{\bB})$ throughout the rest of this paper.
\subsection{Channel Model}
The mmWave MIMO channel is characterized by a standard multi-path model\cite[ch. 7.3.2]{tse2005fundamentals}:
\begin{align}\label{MCM}
\bH\!=\!\sqrt{\frac{N_\mathrm{r}N_\mathrm{t}}{L}}\sum_{\ell=1}^{L}\gamma_{\ell}\ba(\theta_{\mathrm{r},\ell})\ba(\theta_{\mathrm{t},\ell})^H
\end{align}
where $L$ denotes the number of physical propagation paths between the transmitter
and the receiver; $\gamma_\ell$ represents the complex gain of the $\ell$th propagation path; We assume that $\gamma_\ell$ are i.i.d. complex Gaussian distributed with zero-mean and unit-variance; $\ba(\theta_{\mathrm{r},\ell})$ and $\ba(\theta_{\mathrm{t},\ell})$ represent the receive and transmit array steering vectors, with $\theta_{\textrm{r},\ell}$ and $\theta_{\textrm{t},\ell}$ being the angles of arrival (AOA) and the angles of departure (AOD), respectively. In this paper, the transmitter and receiver adopt uniform linear arrays, whose array steering vector $\ba(\theta)$ is given by
\begin{align}
\ba(\theta)\!=\!\frac{1}{\sqrt{N}}\Big[1, e^{-j\frac{2\pi}{\lambda}d\sin\theta},...,e^{-j\frac{2\pi}{\lambda}d(N-1)\sin\theta}\Big]^T
\end{align}
where $N$ is the number of antenna element, $\lambda$ is the wavelength of the carrier frequency and $d\!=\!\frac{1}{2}\lambda$ is the antenna spacing.

The channel model in \eqref{MCM} can be rewritten more compactly as
\begin{align}\label{MCMC}
\bH\!=\!\sqrt{\frac{N_\mathrm{r}N_\mathrm{t}}{L}}\bA_\textrm{r}\bm{\Gamma}\bA_\textrm{t}^{\!H}
\end{align}
where $\bm{\Gamma}\!=\!\diag(\gamma_1,...,\gamma_L)$; $\bA_\textrm{r}\!\in\! \mathds{C}^{N_{\textrm{r}}\times L}$ and $\bA_\textrm{t}\!\in\! \mathds{C}^{N_{\textrm{t}}\times L}$ are stacked array steering vectors of AOA and AOD respectively, given by
\begin{align}
&\bA_\textrm{r}\!=\!\big[\ba(\theta_{\textrm{r},1}),...,\ba(\theta_{\textrm{r},L})\big]\\
&\bA_\textrm{t}\!=\!\big[\ba(\theta_{\textrm{t},1}),...,\ba(\theta_{\textrm{t},L})\big]. \label{TASM}
\end{align}

This work assumes that the small scale fading $\bm{\Gamma}$ varies rapidly while the variation of angle information $\bA_\textrm{r}$ and $\bA_\textrm{t}$ is slow\cite{gao2017fast}. Since the angle information changes slowly, we further assume that the transmitter can obtain statistical CSI through feedback, i.e., the transmitter knows $\bA_\textrm{r}$ and $\bA_\textrm{t}$.

\section{Problem Formulation}
For mmWave MIMO systems, it may not be practical to obtain the instantaneous CSI by conventional channel estimation techniques because 1) the channel matrix measured in the baseband depends on the choice of analog precoder; 2) the training blocks may be prohibitively long due to the large bandwidth and low signal-to-noise ratio (SNR). To mitigate this difficulty, we propose new formulations in which analog and/or digital precoders are designed under statistical CSI.
\subsection{Statistical-CSI-Based Formulation}
We assume that the transmitter has the knowledge of statistical CSI, including $\bA_\mathrm{r}$, $\bA_\mathrm{t}$ and the distribution of $\bm{\Gamma}$. Then we design the analog and digital precoder to maximize the average mutual information. Suppose each entry of the input data vector $\bx$ is uniformly distributed from a given constellation set with cardinality $M$. The average mutual information between $\bx$ and $\by$ is given by
\begin{align}\label{J1}
E_{\bH}\mathcal{I}(\bx;\by|\bH)
\end{align}
where $\mathcal{I}(\bx;\by|\bH)$ is the instantaneous mutual information between $\bx$ and $\by$ \cite{xiao2011globally}
\begin{align}\label{J2}
\mathcal{I}(\bx;\by|\bH)\!=\!\log K
\!-\!\frac{1}{K}\sum_{m=1}^{K} \!E_{\bn}\left\{\log\sum_{k=1}^{K}\exp(-d_{mk})\right\}.
\end{align}
Here $K\!=\!M^{N_\mathrm{s}}$ is a constant number, and $d_{mk}\!=\!\sigma^{-2}\left(\|\bH\bar{\bF}\bar{\bB}(\bx_m\!-\!\bx_k)\!+\!\bn\|^2\!-\!\|\bn\|^2\right)$, with $\bx_m$ and $\bx_k$ being two possible data vectors taken from $\bx$. The average mutual information maximization problem can then be formulated as
\begin{equation}\label{J3}
\begin{aligned}
\underset{\{\mathcal{S}_j\}\in \mathcal{S}}{\mathrm{maximize}} \quad R(\{\mathcal{S}_j\})
\end{aligned}
\end{equation}
where $R(\{\mathcal{S}_j\})$ is the maximum average mutual information with given partition of subsets, i.e.,
\begin{equation}\label{J4}
\begin{aligned}
R(\{\mathcal{S}_j\})\!=\underset{\bar{\bF}\in \mathcal{F},\bar{\bB}\in \mathcal{B}}{\mathrm{maximize}}\quad E_{\bH}\mathcal{I}(\bx;\by|\bH).
\end{aligned}
\end{equation}

Problem \eqref{J3} is a combinatorial optimization problem for which finding the optimal solution requires an exhaustive search over all nonempty $\{\mathcal{S}_j\}_{j=1}^{N_\mathrm{rf}}$ in $\mathcal{S}$. The total number of combinations is known as Stirling number of the second kind \cite{knuth1989concrete} and is given by
\begin{align}
|\mathcal{S}|\!=\!\frac{1}{N_\mathrm{rf}!}\sum_{k=0}^{N_\mathrm{rf}}(-1)^{N_\mathrm{rf}-k}\binom{N_\mathrm{t}}{k}k^{N_\mathrm{t}}.
\end{align}
Then we can rewrite problem \eqref{J3} as
\begin{equation}\label{J5}
\begin{aligned}
\underset{\ell\in\{1,...,|\mathcal{S}|\}}{\mathrm{maximize}}\quad
R(\{\mathcal{S}_{j,\ell}\})
\end{aligned}
\end{equation}
where $\{\mathcal{S}_{j,\ell}\}$ represents the $\ell$th given partition of subsets belonging to $\mathcal{S}$.

Although \eqref{J5} provides a theoretically possible way for solving problem \eqref{J3}, its computational complexity is prohibitive even for a small number of transmit antennas and RF chains. For example, when $N_\mathrm{t}\!=\!16$ and $N_\mathrm{rf}\!=\!4$, $|\mathcal{S}|$ is equal to $1.718\times 10^8$, which implies that we need to solve problem \eqref{J4} over ten million times to obtain the optimal analog and digital precoder.

We propose a new formulation to reduce the computational complexity of problem \eqref{J3}. Recall that $\{\mathcal{S}_j\}_{j=1}^{N_\mathrm{rf}}$ represent positions of nonzero entries in $\bar{\bF}$, and the role of $\bar{\bF}$ is to reshape the effective channel matrix $\bH\bar{\bF}$. Therefore, we design $\{\mathcal{S}_j\}_{j=1}^{N_\mathrm{rf}}$ and the corresponding $\bar{\bF}$ such that the average effective channel gain $E_{\bH}\|\bH\bar{\bF}\|_F^2$ is maximized. The dynamic subarray design problem can then be formulated as
\begin{equation}\label{J6}
\begin{aligned}
\underset{\bar{\bF}\in \mathcal{F},\{\mathcal{S}_j\}\in \mathcal{S}}{\mathrm{maximize}}\quad E_{\bH}\trace\left(\bar{\bF}^H\bH^H\bH\bar{\bF}\right).
\end{aligned}
\end{equation}

We solve problem \eqref{J6} to obtain its optimal solutions, denoted by $\bar{\bF}_{\!\mathrm{init}}^\star$ and $\{\mathcal{S}_j^\star\}_{j=1}^{N_\mathrm{rf}}$. Then we solve problem \eqref{J4} with given $\{\mathcal{S}_j^\star\}_{j=1}^{N_\mathrm{rf}}$ to obtain the optimally effective analog and digital precoders $(\bar{\bF}^\star,\bar{\bB}^\star)$. Note that since $\bar{\bF}_{\!\mathrm{init}}^\star$ is not obtained by maximizing the average mutual information, we do not use it directly as the optimally effective analog precoder. However, the solution $\bar{\bF}_{\!\mathrm{init}}^\star$ serves as a good initial point for solving problem \eqref{J4}. Therefore, we first design a low complexity algorithm to solve problem \eqref{J6}, and then design an effective algorithm to solve the hybrid precoding problem \eqref{J4} with given $\{\mathcal{S}_j^\star\}_{j=1}^{N_\mathrm{rf}}$.

\subsection{Mixed-CSI-Based Formulation}
The basic idea of mixed CSI based formulation is to design the analog precoder based on statistical CSI, and then estimate the reduced-dimensional effective channel matrix $\bH\bar{\bF}^\star$, where $\bar{\bF}^\star\!=\!\bF(\bF^H\bF)^{-\frac{1}{2}}$ is the optimally effective analog precoder based on statistical CSI. After that, the transmitter utilizes the \emph{instantaneous} effective channel matrix $\bH\bar{\bF}^\star$ to design effective digital precoder $\bar{\bB}$, and this is a typical MIMO precoding problem. In this case, the burden of channel estimation is greatly reduced because the dimension of $\bH\bar{\bF}^\star\!\in\!\mathds{C}^{N_\mathrm{r}\times N_\mathrm{rf}}$ is much smaller than that of $\bH\!\in\!\mathds{C}^{N_\mathrm{r}\times N_\mathrm{t}}$.

Given the instantaneous effective channel matrix $\bH\bar{\bF}$, the digital precoding problem can be expressed by
\begin{equation}\label{J7}
\begin{aligned}
C(\bH\bar{\bF})\!=\!\underset{\bar{\bB}\in \mathcal{B}}{\mathrm{maximize}}\quad\mathcal{I}(\bx;\by|\bH)
\end{aligned}
\end{equation}
where $C(\bH\bar{\bF})$ is the maximum mutual information under the given effective channel matrix $\bH\bar{\bF}$. Then the mixed-CSI-based hybrid precoding problem can be formulated as
\begin{equation}\label{J8}
\begin{aligned}
\underset{\bar{\bF}\in \mathcal{F},\{\mathcal{S}_j\}\in \mathcal{S}}{\mathrm{maximize}}\quad E_{\bH}C(\bH\bar{\bF}).
\end{aligned}
\end{equation}
Problem \eqref{J8} is intractable because it is prohibitive to compute the objective function $E_{\bH}C(\bH\bar{\bF})$. In order to estimate $E_{\bH}C(\bH\bar{\bF})$ at a given point $\bar{\bF}$, we need to solve the nonconvex problem \eqref{J7} thousands of times for randomly generated channel matrix $\bH$. To mitigate this difficulty, we replace $E_{\bH}C(\bH\bar{\bF})$ by a computationally efficient bound. Invoke Jensen's inequality, $E_{\bH}C(\bH\bar{\bF})$ can be lower bounded by
\begin{equation}\label{J9}
\begin{aligned}
E_{\bH}C(\bH\bar{\bF})\!\geq\underset{\bar{\bB}\in \mathcal{B}}{\mathrm{maximize}}\quad E_{\bH}\mathcal{I}(\bx;\by|\bH).
\end{aligned}
\end{equation}
Replacing $E_{\bH}C(\bH\bar{\bF})$ by its lower bound, problem \eqref{J8} is approximated as
\begin{equation}\label{J10}
\begin{aligned}
& \underset{\bar{\bF}\in \mathcal{F},\{\mathcal{S}_j\}\in \mathcal{S},\bar{\bB}\in \mathcal{B}}{\mathrm{maximize}}\quad E_{\bH}\mathcal{I}(\bx;\by|\bH)
\end{aligned}
\end{equation}
which is exactly the same as problem \eqref{J3}. Then we can use the same procedure to solve this problem, i.e., we first solve problem \eqref{J6} to obtain $\{\mathcal{S}_j^\star\}_{j=1}^{N_\mathrm{rf}}$, and then solve problem \eqref{J4} with given $\{\mathcal{S}_j^\star\}_{j=1}^{N_\mathrm{rf}}$ to obtain the optimally effective analog precoder. Note that although the statistical-CSI-based formulation and the mixed-CSI-based formulation solve the same optimization problem, there is an important difference between them. The optimization variable $\bar{\bB}$ in the mixed-CSI-based formulation is just an auxiliary variable made for analog precoder design. After obtaining the optimally effective analog precoder, the real digital precoder should be obtained by solving problem \eqref{J7}.

\section{Dynamic Subarray Design}
In this section, we propose a low complexity algorithm to solve problem \eqref{J6}. Note that the objective function in problem \eqref{J6} can be rewritten as
\begin{align}\label{K0}
E_{\bH}&\trace\left(\bar{\bF}^H\bH^H\bH\bar{\bF}\right)\nonumber\\
&=\frac{N_\mathrm{r}N_\mathrm{t}}{L}E_{\bm{\Gamma}}\trace\left(\bar{\bF}^H\bA_\textrm{t}\bm{\Gamma}^{H}\bA_\textrm{r}^H\bA_\textrm{r}\bm{\Gamma}\bA_\textrm{t}^{\!H}\bar{\bF}\right)\nonumber\\
&=\frac{N_\mathrm{r}N_\mathrm{t}}{L} \trace\left(\bar{\bF}^H\bA_\textrm{t}\bA_\textrm{t}^{\!H}\bar{\bF}\right)
\end{align}
where the second equality in equation \eqref{K0} holds because $E_{\bm{\Gamma}}(\bm{\Gamma}^{H}\bA_\textrm{r}^H\bA_\textrm{r}\bm{\Gamma})\!=\!\bI$.
Plugging $\bar{\bF}\!=\!\bF(\bF^H\bF)^{-\frac{1}{2}}$ into equation \eqref{K1}, we obtain the following problem
\begin{equation}\label{K1}
\begin{aligned}
& \underset{\bF, \{\mathcal{S}_j\}}{\mathrm{maximize}}
&& \trace\Big[(\bF^H\bF)^{-\frac{1}{2}}\bF^H\bA_\textrm{t}\bA_\textrm{t}^{\!H}\bF(\bF^H\bF)^{-\frac{1}{2}}\Big]\\
& \mathrm{subject \;to}
& &|\bF_{\!ij}|\!=\!\mathbf{1}_{\mathcal{S}_j}(i),\forall (i,j)\\
&&&\{\mathcal{S}_j\}\in \mathcal{S}.
\end{aligned}
\end{equation}
It is difficult to solve problem \eqref{K1} directly because the feasible set of problem \eqref{K1} is characterized by $\bF$ and $\{\mathcal{S}_j\}_{j=1}^{N_\mathrm{rf}}$. To address this issue, the following proposition rewrites the feasible set as explicit constraints of $\bF$.
\begin{proposition}
The feasible set of problem \eqref{K1} can be expressed by
\begin{align}\label{K2}
\begin{aligned}
&|\bF_{\!ij}|\!\in\!\{0,1\},\;\; \forall (i,j)\\
&\|\bF_{\!i\bullet}\|_0\!=\!1,\;\; \forall i
\end{aligned}
\end{align}
where $\bF_{\!i\bullet}$ denotes the $i$th row of $\bF$, and $\|\cdot\|_0$ represents the total number of nonzero elements in a vector.
\end{proposition}
\begin{IEEEproof}
See Appendix.
\end{IEEEproof}

According to Proposition 1, we rewrite problem \eqref{K1} as
\begin{equation}\label{K3}
\begin{aligned}
& \underset{\bF}{\mathrm{maximize}}
& &\trace\left[(\bF^H\bF)^{-\frac{1}{2}}\bF^H\bA_\mathrm{t}\bA_\mathrm{t}^H\bF(\bF^H\bF)^{-\frac{1}{2}}\right]\\
& \mathrm{subject \;to}
& &|\bF_{\!ij}|\!\in\!\{0,1\},\;\; \forall (i,j)\\
&&&\|\bF_{\!i\bullet}\|_0\!=\!1,\;\; \forall i.
\end{aligned}
\end{equation}
Problem \eqref{K3} is still intractable due to nonconvex discrete constraints $|\bF_{\!ij}|\!\in\!\{0,1\}$ and $\|\bF_{\!i\bullet}\|_0\!=\!1$. Therefore, we first drop the constraints and consider the unconstrained problem
\begin{equation}\label{K4}
\begin{aligned}
& \underset{\bF}{\mathrm{maximize}}\quad
\trace\left[(\bF^H\bF)^{-\frac{1}{2}}\bF^H\bA_\textrm{t}\bA_\textrm{t}^{\!H}\bF(\bF^H\bF)^{-\frac{1}{2}}\right].
\end{aligned}
\end{equation}
Problem \eqref{K4} is a generalized eigenvalue problem, and its optimal solution is given by\cite{park2017dynamic}
\begin{align}
\bF\!=\!\bU_{\!\bA}\bR
\end{align}
where $\bU_{\!\bA}\!\in\!\mathds{C}^{N_\mathrm{t}\times N_\mathrm{rf}}$ is the left singular vectors of $\bA_\textrm{t}$ corresponding to the largest $N_\mathrm{rf}$ singular values, and $\bR\!\in\!\mathds{C}^{N_\mathrm{rf}\times N_\mathrm{rf}}$ is an arbitrary unitary matrix. Note that when $L<N_\mathrm{rf}$, the remaining $N_\mathrm{rf}-L$ left singular vectors in $\bU_{\!\bA}$ can be chosen arbitrarily as long as $\bU_{\!\bA}$ satisfies $\bU_{\!\bA}^H\bU_{\!\bA}\!=\!\bI$.

In general, if there exists a unitary matrix $\bR$ such that the unconstrained optimal solution $\bU_{\!\bA}\bR$ satisfies \eqref{K2}, then $\bU_{\!\bA}\bR$ is the globally optimal solution of problem \eqref{K3}. However, such $\bR$ may not exist and thus we use $\bU_{\!\bA}\bR$ to find a nearby feasible solution. Specifically, consider the following optimization problem
\begin{equation}\label{K6}
\begin{aligned}
& \underset{\bF,\bR\in \mathcal{U}}{\mathrm{minimize}}
& &\|\bF\!-\!\bU_{\!\bA}\bR\|_F^2\\
& \mathrm{subject \;to}
& &|\bF_{\!ij}|\!\in\!\{0,1\},\;\; \forall (i,j)\\
&&&\|\bF_{\!i\bullet}\|_0\!=\!1,\;\; \forall i
\end{aligned}
\end{equation}
where $\mathcal{U}$ denotes the set of unitary matrices. Since the optimization variables $\bF$ and $\bR$ are separate, we adopt the alternating minimization approach to solve problem \eqref{K6}.

Given $\bR$, the optimal $\bF$ of problem \eqref{K6} has a simple closed form solution. Let $j^\star(i)\!=\!\mathrm{argmax}_{1\leq j\leq N_\mathrm{rf}}\left|[\bU_{\!\bA}\bR]_{ij}\right|$,  then the optimal $\bF$ of problem \eqref{K6} can be expressed by
\begin{align}\label{K9}
\bF_{\!ij}\!=\!\left\{
\begin{aligned}
&\frac{[\bU_{\!\bA}\bR]_{ij}}{\left|[\bU_{\!\bA}\bR]_{ij}\right|}\quad\quad \mathrm{if}\;\;j\!=\!j^\star(i)\\
&\;\;0\quad\quad\quad\quad\quad\;\; \mathrm{otherwise}.
\end{aligned}
\right.
\end{align}

Given $\bF$, problem \eqref{K6} is reduced to an orthogonal procrustes problem
\begin{equation}\label{K10}
\begin{aligned}
& \underset{\bR\in \mathcal{U}}{\mathrm{minimize}}\quad\|\bF\!-\bU_{\!\bA}\bR\|_F^2.
\end{aligned}
\end{equation}
Let the singular value decomposition of $\bZ\!=\!\bF^H\bU_{\!\bA}$ be
\begin{align}
\bZ\!=\!\bF^H\bU_{\!\bA}\!=\!\bU_{\bZ}\pmb{\Sigma}_{\bZ}\bV_{\!\bZ}^H
\end{align}
where $\bU_{\bZ}$ is a unitary matrix with left singular vectors, $\pmb{\Sigma}_{\bZ}$ is a diagonal matrix with singular values arranged in decreasing order, and $\bV_{\!\bZ}$ is another unitary matrix with right singular vectors. Then the optimal solution of problem \eqref{K10} is given by \cite{horn1994matrix}
\begin{align}\label{K11}
\bR\!=\!\bV_{\!\bZ}\bU_{\bZ}^H.
\end{align}
Combining \eqref{K9} and \eqref{K11}, we propose a simple alternating minimization algorithm to solve problem \eqref{K6} and obtain the corresponding near optimal partition of subsets $\{\mathcal{S}_j\}_{j=1}^{N_\mathrm{rf}}$. The details of this algorithm is summarized in Algorithm 1.
\begin{algorithm}
\caption{Dynamic subarray design}
\begin{algorithmic}
\STATE 1. Given the stacked array steering vectors of AOD $\bA_\mathrm{t}$. Compute $\bA_\mathrm{t}$'s left singular vectors $\bU_{\!\bA}$ and generate an arbitrary initial unitary matrix ${\bR}$.
\STATE 2. \textbf{While} the stopping criterion is not satisfied
\begin{itemize}
  \item Given $\bR$, solve problem \eqref{K6} to obtain the optimal $\bF$ in \eqref{K9}.
  \item Given $\bF$, solve problem \eqref{K6} to obtain the optimal $\bR$ in \eqref{K11}.
\end{itemize}
\STATE 3. Return $\bar{\bF}_{\!\mathrm{init}}^\star\!=\!\bF(\bF^H\bF)^{-\frac{1}{2}}$ and the corresponding $\{\mathcal{S}_j^\star\}$.
\end{algorithmic}
\end{algorithm}

We conclude this section with several remarks on Algorithm 1:
\begin{itemize}
  \item The convergence of Algorithm 1 is guaranteed because the objective function $\|\bF\!-\!\bU_{\!\bA}\bR\|_F^2$ is bounded, and it is decreasing in each iteration.
  \item Since problem \eqref{K6} is a nonconvex problem, the solution obtained by Algorithm 1 depends on the initial unitary matrix $\bR$. Therefore, we can run Algorithm 1 several times with different initial $\bR$, and then choose the solution corresponding to the largest $\left\|\bA_\textrm{t}^{\!H}\bar{\bF}^\star_{\!\mathrm{init}}\right\|_F^2$.
  \item When $\bar{\bF}_{\!\mathrm{init}}^\star\!=\!\bF(\bF^H\bF)^{-\frac{1}{2}}$ is determined, the corresponding $\{\mathcal{S}_j^\star\}$ is given by
\begin{align*}
\mathcal{S}_j^\star\!=\!\left\{i\Big|\left|[\bar{\bF}_{\!\mathrm{init}}^\star]_{ij}\right|\!\neq\! 0\right\},\;j\!=\!1,...,N_{\mathrm{rf}}.
\end{align*}
\end{itemize}

\section{Hybrid Precoding With Finite-Alphabet Inputs}
In this section, we first derive the lower bound for the average mutual information $E_{\bH}\mathcal{I}(\bx;\by|\bH)$, and then propose an effective algorithm to design analog and digital precoders.
\subsection{Lower Bound For Average Mutual Information}
It is difficult to compute and optimize the average constellation-constrained mutual information directly because both $E_{\bH}\mathcal{I}(\bx;\by|\bH)$ and its gradient have no closed form expressions. To estimate $E_{\bH}\mathcal{I}(\bx;\by|\bH)$ as well as its gradient, we need to use Monte Carlo method and/or numerical integral, whose computational complexity are prohibitively high.

This difficulty can be partially mitigated by the following proposition, which provides the lower bound of $E_{\bH}\mathcal{I}(\bx;\by|\bH)$ in closed form.
\begin{proposition}
The average constellation-constrained mutual information of mmWave MIMO channels can be lower bounded by
\begin{align}
L(\bar{\bF},\bar{\bB})\!=\!\log K&\!-\!N_\textrm{r}\left(\frac{1}{\ln 2}\!-\!1\right)\!-\!\frac{1}{K}\sum_{m=1}^{K}\nonumber\\
&\log\sum_{k=1}^K\det\left[\bI\!+\!\left(\bA_\textrm{r}^H\bA_\textrm{r}\right)^T\!\circ\!\bW_{\!mk}\right]^{-1}
\end{align}
where
\begin{align}
\bW_{\!mk}\!=\!\frac{N_\mathrm{r}N_\mathrm{t}}{2\sigma^2L}\bA_\textrm{t}^H\bar{\bF}\bar{\bB}(\bx_m\!-\!\bx_k)(\bx_m\!-\!\bx_k)^H\bar{\bB}^H\bar{\bF}^H\bA_\textrm{t}.
\end{align}
\end{proposition}
\begin{IEEEproof}
See Appendix.
\end{IEEEproof}

The computational complexity of the lower bound $L(\bar{\bF},\bar{\bB})$ is still very high because it needs to calculate the determinant $K^2$ times. For example, when we adopt 16QAM modulation ($M\!=\!16$) and the number of data streams $N_\mathrm{s}$ is 4, $K^2$ is equal to $4.295\times 10^9$. To further reduce the complexity, we notice that the receive steering vectors are asymptotically orthogonal to each other when the number of receive antennas $N_\textrm{r}$ approaches infinity, i.e., $\lim_{N_\mathrm{r}\rightarrow \infty}\bA_\textrm{r}^H\bA_\textrm{r}\!=\!\bI$. Based on this observation, we derive a low complexity approximation of $L(\bar{\bF},\bar{\bB})$ in the following proposition.
\begin{proposition}
The lower bound $L(\bar{\bF},\bar{\bB})$ can be approximated by
\begin{align}
L_{\!A}(\bar{\bF},\bar{\bB})\!=\!\log K&\!-\!N_\textrm{r}\left(\frac{1}{\ln 2}\!-\!1\right)\!-\!\frac{1}{K}\sum_{m=1}^{K}\nonumber\\
&\log\sum_{k=1}^K\prod_{\ell=1}^{L}\left(1\!+\!\frac{N_\mathrm{r}N_\mathrm{t}}{2\sigma^2L}\left|\beta_{mk\ell}\right|^2\right)^{-1}
\end{align}
where $\beta_{mk\ell}\!=\!\ba(\theta_{\textrm{t},\ell})^H\bar{\bF}\bar{\bB}(\bx_m\!-\!\bx_k)$, with $\ba(\theta_{\textrm{t},\ell})$ being the $\ell$th column of $\bA_\textrm{t}$. In addition, the limit of $L_{\!A}(\bar{\bF},\bar{\bB})$ is $L(\bar{\bF},\bar{\bB})$ as $N_\mathrm{r}$ approaches infinity.
\end{proposition}
\begin{IEEEproof}
See Appendix.
\end{IEEEproof}

The accuracy and computational complexity of the lower bound and its approximation will be shown in Fig. 1 and Table 1 in the simulation result section.

\subsection{Hybrid Precoding Design}
In this section, we solve the hybrid precoding problem \eqref{J4} with given $\{\mathcal{S}_j^\star\}_{j=1}^{N_\mathrm{rf}}$ obtained by Algorithm 1. First, by replacing the average mutual information $E_{\bH}\mathcal{I}(\bx;\by|\bH)$ with the approximated lower bound $L_{\!A}(\bar{\bF},\bar{\bB})$, problem \eqref{J4} can be approximated as
\begin{equation}\label{L2}
\begin{aligned}
&\underset{\bar{\bF},\bar{\bB}}{\mathrm{maximize}}
&& L_{\!A}(\bar{\bF},\bar{\bB})\\
& \mathrm{subject \;to}
& &\left|\bar{\bF}_{\!ij}\right|\!=\!\left|\mathcal{S}_j^\star\right|^{-\frac{1}{2}}\mathbf{1}_{\mathcal{S}_j}(i),\;\forall (i,j)\\
&&& \trace\left(\bar{\bB}^H\bar{\bB}\right)\!\leq\! P.
\end{aligned}
\end{equation}
Note that the constraint $|\bar{\bF}_{\!ij}|\!=\!|\mathcal{S}_j^\star|^{-\frac{1}{2}}\mathbf{1}_{\mathcal{S}_j}(i)$ implies that only the phase of nonzero $|\bar{\bF}_{\!ij}|$ can be changed. Therefore, instead of using $\bar{\bF}$ as the optimization variable, it is more convenient to optimize the phase of nonzero entries in $\bar{\bF}$. Define the phase matrix $\pmb{\Phi}$ as
\begin{align}
\pmb{\Phi}_{\!ij}\!=\!\angle \bar{\bF}_{\!ij}\mathbf{1}_{\mathcal{S}_j^\star}(i),\;\forall (i,j)
\end{align}
where $\angle \bar{\bF}_{\!ij}$ represents the phase of $\bar{\bF}_{\!ij}$. Then $\bar{\bF}$ can be expressed as
\begin{align}
\bar{\bF}_{\!ij}\!=\!\left|\mathcal{S}_j^\star\right|^{-\frac{1}{2}}\exp(\jmath\pmb{\Phi}_{\!ij})\mathbf{1}_{\mathcal{S}_j^\star}(i),\;\forall (i,j).
\end{align}
Using $\pmb{\Phi}$ as the optimization variable and defining a new function $R(\pmb{\Phi},\bar{\bB})\!\triangleq\!L_{\!A}(\bar{\bF}(\pmb{\Phi}),\bar{\bB})$, problem \eqref{L2} can be rewritten as
\begin{equation}\label{L3}
\begin{aligned}
&\underset{\pmb{\Phi},\bar{\bB}}{\mathrm{maximize}}
&& R(\pmb{\Phi},\bar{\bB})\\
& \mathrm{subject \;to}
&& \trace\left(\bar{\bB}^H\bar{\bB}\right)\!=\! P.
\end{aligned}
\end{equation}
Here we express the power constraint as $\trace\left(\bar{\bB}^H\bar{\bB}\right)\!=\! P$ because $R(\pmb{\Phi},\bar{\bB})$ is monotonically increasing with respect to $\|\bar{\bB}\|_F^2$. Then we provide the gradient of $R(\pmb{\Phi},\bar{\bB})$ in the following proposition, which forms the foundation for solving problem \eqref{L3}.
\begin{proposition}
The gradient of $R(\pmb{\Phi},\bar{\bB})$ with respect to $\bar{\bB}$ and $\pmb{\Phi}$ are given by
\begin{align}\label{L31}
&\nabla_{\!\bar{\bB}}R(\pmb{\Phi},\bar{\bB})\!=\!\sum_{\ell=1}^L \bar{\bF}^H\ba(\theta_{\textrm{t},\ell})\ba(\theta_{\textrm{t},\ell})^H\bar{\bF}\bar{\bB}\bE_{\ell} \nonumber\\
&\nabla_{\!\pmb{\Phi}}R(\pmb{\Phi},\bar{\bB})\!=\!2\sum_{\ell=1}^L \Im\left[\bar{\bF}^H\ba(\theta_{\textrm{t},\ell})\ba(\theta_{\textrm{t},\ell})^H\bar{\bF}\bar{\bB}\bE_{\ell}\bar{\bB}^H\!\circ\! \bar{\bF}^*\right]
\end{align}
where
\begin{align}
\bE_{\ell}\!=\!\frac{1}{\ln(2)\!\cdot\!K}\sum_{m,k}\zeta_{mk\ell}(\bx_m\!-\!\bx_k)(\bx_m\!-\!\bx_k)^H
\end{align}
with
\begin{align*}
\zeta_{mk\ell}\!=&\!\left(\frac{2\sigma^2L}{N_\mathrm{r}N_\mathrm{t}}\!+\!|\beta_{mk\ell}|^2\right)^{-1}\!\!\cdot\!\prod_{\ell=1}^{L}\left(1\!+\!\frac{N_\mathrm{r}N_\mathrm{t}}{2\sigma^2L}\left|\beta_{mk\ell}\right|^2\right)^{-1}\!\cdot\!\\
&\left[\sum_{k=1}^K\prod_{\ell=1}^{L}\left(1\!+\!\frac{N_\mathrm{r}N_\mathrm{t}}{2\sigma^2L}\left|\beta_{mk\ell}\right|^2\right)^{-1}\right]^{-1}.
\end{align*}
\end{proposition}
\begin{IEEEproof}
See Appendix.
\end{IEEEproof}

We propose a manifold-based gradient ascent algorithm to optimize $\pmb{\Phi}$ and $\bar{\bB}$ simultaneously using the gradient information. At the $k$th iteration, the algorithm updates the current solution $(\pmb{\Phi}_{\!k},\bar{\bB}_k)$ to $(\pmb{\Phi}_{\!k+1},\bar{\bB}_{k+1})$ by the following rules
\begin{align}\label{L7}
&\pmb{\Phi}_{\!k+1}\!=\!\pmb{\Phi}_{\!k}\!+\!\rho_k\nabla_{\!\pmb{\Phi}}R(\pmb{\Phi}_{\!k},\bar{\bB}_{k})\nonumber\\
&\bar{\bB}_{k+1}\!=\!\mathrm{Proj}\left[\bar{\bB}_k\!+\!\rho_k\mathrm{grad}_{\bar{\bB}}R(\pmb{\Phi}_{\!k},\bar{\bB}_{k})\right]
\end{align}
where $\rho_k>0$ is the stepsize, $\mathrm{Proj}\left[\bar{\bB}_k\right]\!=\!P^{\frac{1}{2}}\|\bar{\bB}_k\|_F^{-1}\bar{\bB}_k$, and $\mathrm{grad}_{\bar{\bB}}R(\pmb{\Phi},\bar{\bB})$ is the gradient of $R(\pmb{\Phi},\bar{\bB})$ on the following (sphere) manifold
\begin{align}\label{L8}
\mathcal{M}\!=\!\left\{\bar{\bB}\big|\trace\left(\bar{\bB}^H\bar{\bB}\right)\!=\! P\right\}.
\end{align}
Based on the definition, $\mathrm{grad}_{\bar{\bB}}R(\pmb{\Phi},\bar{\bB})$ can be computed by projecting $\nabla_{\!\bar{\bB}}R(\pmb{\Phi},\bar{\bB})$ onto the tangent space $T_{\bar{\bB}}\mathcal{M}$ at $\bar{\bB}$, where $T_{\bar{\bB}}\mathcal{M}$ is given by
\begin{align}\label{L9}
T_{\bar{\bB}}\mathcal{M}\!=\!\left\{\bar{\bX}\big|\trace\left(\bar{\bX}^H\bar{\bB}\!+\!\bar{\bB}^H\bar{\bX}\right)\!=\!0\right\}.
\end{align}
Then $\mathrm{grad}_{\bar{\bB}}R(\pmb{\Phi},\bar{\bB})$ can be expressed by
\begin{align}\label{L10}
\mathrm{grad}_{\bar{\bB}}R(\pmb{\Phi},\bar{\bB})\!=\!\underset{\bar{\bX}\in T_{\bar{\bB}}\mathcal{M}}{\mathrm{argmax}} \;\; \left\|\bar{\bX}\!-\!\nabla_{\!\bar{\bB}}R\right\|_F^2.
\end{align}
Using the standard Lagrangian multiplier method, the closed form solution of problem \eqref{L10} is given by
\begin{align}\label{L11}
\mathrm{grad}_{\bar{\bB}}R(\pmb{\Phi},\bar{\bB})\!=\!\nabla_{\!\bar{\bB}}R(\pmb{\Phi},\bar{\bB})\!-\!\frac{\Re\trace\left[(\nabla_{\!\bar{\bB}}R)^H\bar{\bB}\right]}{P}\bar{\bB}.
\end{align}

After obtaining the ascent direction, we need to determine the stepsize $\rho_k$ such that the objective function $R(\pmb{\Phi},\bar{\bB})$ is increasing in each iteration. We propose a modified backtracking line search method, which is usually more efficient than the classic backtracking line search \cite{boyd2004convex}. The main idea is to use $\rho_{k-1}$ as the initial guess of $\rho_k$, and then either increases or decreases it to find the largest $\rho_k$ such that
\begin{align*}
f(\rho_k)\!\triangleq& R(\pmb{\Phi}_{\!k+1},\bar{\bB}_{k+1})\!-\!R(\pmb{\Phi}_{\!k},\bar{\bB}_{k})\!-\!\rho_k\beta_{\mathrm{ga}}\\
&\big(\|\nabla_{\!\pmb{\Phi}}R(\pmb{\Phi}_{\!k},\bar{\bB}_{k})\|_F^2\!+\!\|\mathrm{grad}_{\bar{\bB}}R(\pmb{\Phi}_{\!k},\bar{\bB}_{k})\|_F^2\big)\!\geq\!0
\end{align*}
where $\beta_{\mathrm{ga}}\!\in\![0,0.5]$ is a constant to control the stepsize. Specifically, the stepsize $\rho_k$ is set as
\begin{align}\label{L12}
\rho_k\!=\!\left\{
\begin{aligned}
&2^{K_1-1}\!\cdot\!\rho_{k-1} \quad\quad \mathrm{if}\; f(\rho_{k-1})\!\geq\!0\\
&\!\Big(\frac{1}{2}\Big)^{K_2}\!\!\cdot\!\rho_{k-1} \quad\quad \mathrm{if}\; f(\rho_{k-1})\!<\!0
\end{aligned}
\right.
\end{align}
where $K_1\!\geq\!0$ is the smallest integer such that $f(2^{K_1}\rho_{k-1})\!<\!0$, and $K_2\!\geq\!0$ is the smallest integer such that $f([\frac{1}{2}]^{K_2}\rho_{k-1})\!\geq\!0$. The details of our proposed manifold-based gradient ascent algorithm is summarized in Algorithm 2.
\begin{algorithm}
\caption{Manifold-based gradient ascent algorithm}
\begin{algorithmic}
\STATE 1. Given $\{\mathcal{S}_j^\star\}_{j=1}^{N_\mathrm{rf}}$ (obtained by Algorithm 1), $\mathbf{\Phi}_{\!0}$ and $\bar{\bB}_0$. Set $\rho_0\!=\!2$, $\beta_{\mathrm{ga}}\!=\!0.4$, and $\epsilon\!=\!10^{-4}$.
\STATE 2. For $k=0,1,2,...$ (outer iterations)
\begin{itemize}
  \item Compute the gradient of $R(\pmb{\Phi},\bar{\bB})$ with repsect to $\pmb{\Phi}$ and $\bar{\bB}$ at $(\pmb{\Phi}_{\!k},\bar{\bB}_{k})$ by \eqref{L31}. Then use $\nabla_{\!\bar{\bB}}R(\pmb{\Phi}_{\!k},\bar{\bB}_{k})$ to compute the gradient of $R(\pmb{\Phi},\bar{\bB})$ at $\bar{\bB}_{k}$ on the shere manifold by \eqref{L11}.
    \item If $\|\nabla_{\!\pmb{\Phi}}R(\pmb{\Phi}_{\!k},\bar{\bB}_{k})\|_F^2\!+\!\|\mathrm{grad}_{\bar{\bB}}R(\pmb{\Phi}_{\!k},\bar{\bB}_{k})\|_F^2\!<\!\epsilon$, stop.
  \item Utilize the modified backtracking line search to compute the stepsize $\rho_k$ via \eqref{L12}.
  \item Update $(\mathbf{\Phi}_{\!k},\bar{\bB}_k)$ to $(\mathbf{\Phi}_{\!k+1},\bar{\bB}_{k+1})$ by
  \begin{align*}
&\pmb{\Phi}_{\!k+1}\!=\!\pmb{\Phi}_{\!k}\!+\!\rho_k\nabla_{\!\pmb{\Phi}}R(\pmb{\Phi}_{\!k},\bar{\bB}_{k})\\&\bar{\bB}_{k+1}\!=\!\mathrm{Proj}\left[\bar{\bB}_k\!+\!\rho_k\mathrm{grad}_{\bar{\bB}}R(\pmb{\Phi}_{\!k},\bar{\bB}_{k})\right].
\end{align*}
\end{itemize}
\end{algorithmic}
\end{algorithm}

\section{Simulation Results}
We provide several examples in this section to illustrate the relationship and the computational complexity comparison between average mutual information and its lower bound as well as the lower bound approximation. We also show the convergence
of the proposed hybrid precoding algorithm and the efficacy of the designed hybrid precoders. For convenience, we rewrite the angles of arrival $\{\theta_{\textrm{r},\ell}\}_{\ell=1}^L$ as a vector $\bm{\theta}_\mathrm{r}$, whose $\ell$th element corresponds to $\theta_{\textrm{r},\ell}$. Similarly, the angles of departure $\{\theta_{\textrm{t},\ell}\}_{\ell=1}^L$ can be expressed by $\bm{\theta}_\mathrm{t}$. The angles of arrival follow the Laplacian distribution with a fixed or uniformly distributed mean angle $\bar{\theta}_\mathrm{r}$, and a constant angular spread (standard deviation) of $\frac{\pi}{18}$. The angles of departure follow the Laplacian distribution with a fixed mean angle $\bar{\theta}_\mathrm{t}$, and an angular spread of $\frac{\pi}{18}$.

\begin{figure}[!htp]
  \begin{center}
  \includegraphics[scale=0.6]{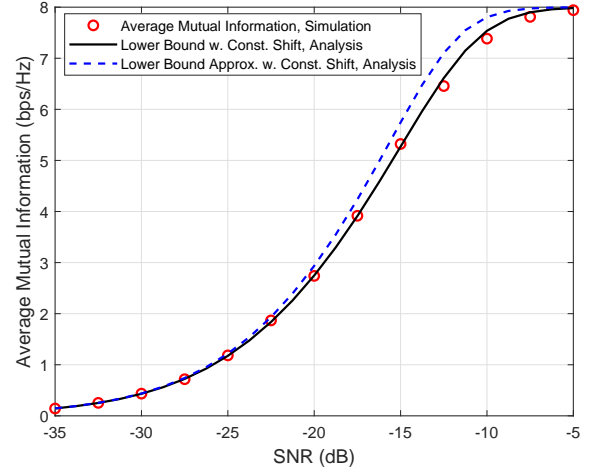}
  \vspace{-0.1cm}
  \caption{Average mutual information with QPSK inputs for mmWave MIMO channels ($N_\mathrm{r}\!=\!32$, $N_\mathrm{t}\!=\!64$, $N_\mathrm{rf}\!=\!4$, $N_\mathrm{s}\!=\!4$, $L=6$).}
\end{center}
  \vspace{-0.5cm}
\end{figure}
\subsection{Example 1: Average Mutual Information and Lower Bound}
This example is utilized to show that 1) the lower bound $L(\bar{\bF},\bar{\bB})$ plus a constant is a very accurate approximation to the average mutual information; 2) the lower bound approximation $L_A(\bar{\bF},\bar{\bB})$ plus a constant is also a good approximation to the average mutual information; 3) the computational complexity of $L_{\!A}(\bar{\bF},\bar{\bB})$ is a few orders of magnitudes lower than that of the average mutual information.

We begin with the consideration of limits of the average mutual information. When the noise power $\sigma^2$ approaches $0$ and $+\infty$, the limits are given by
\begin{align}
&\lim_{\sigma^2\rightarrow 0}E_{\bH}\mathcal{I}(\bx;\by|\bH)\!=\!\log K\\
&\lim_{\sigma^2\rightarrow +\infty}E_{\bH}\mathcal{I}(\bx;\by|\bH)\!=\!0.
\end{align}
At the same time, the limits of $L(\bar{\bF},\bar{\bB})$ are given by
\begin{align}
&\lim_{\sigma^2\rightarrow 0}L(\bar{\bF},\bar{\bB})\!=\!\log K-N_\mathrm{r}\left(\frac{1}{\ln(2)}-1\right)\\
&\lim_{\sigma^2\rightarrow +\infty}L(\bar{\bF},\bar{\bB})\!=\!-N_\mathrm{r}\left(\frac{1}{\ln(2)}-1\right)
\end{align}
which imply that a constant gap $N_\mathrm{r}\Big(\frac{1}{\ln(2)}-1\Big)$ exists between the average mutual information $E_{\bH}\mathcal{I}(\bx;\by|\bH)$ and its lower bound $L(\bar{\bF},\bar{\bB})$ at low and high SNR regimes. Similarly, the same constant gap $N_\mathrm{r}\Big(\frac{1}{\ln(2)}-1\Big)$ exists between $E_{\bH}\mathcal{I}(\bx;\by|\bH)$ and $L_{\!A}(\bar{\bF},\bar{\bB})$. Since the optimized hybrid precoders will remain unchanged by adding a constant value to the objective function, we demonstrate that the lower bound $L(\bar{\bF},\bar{\bB})$ and its approximation $L_{\!A}(\bar{\bF},\bar{\bB})$ plus a constant serve as good approximations to the average mutual information.

We consider a mmWave MIMO system with $N_\mathrm{r}\!=\!32$, $N_\mathrm{t}\!=\!64$, $N_\mathrm{rf}\!=\!4$ and $N_\mathrm{s}\!=\!4$. The number of physical propagation paths is set as $L=6$, and the SNR is defined as $\mathrm{SNR}\!=\!\frac{P}{\sigma^2}$. The input signal is drawn from QPSK modulation. The mean angles of $\bm{\theta_{\textrm{r}}}$ and $\bm{\theta_{\textrm{t}}}$ are set as $\bar{\theta}_\mathrm{r}\!=\!\bar{\theta}_\mathrm{t}\!=\!\frac{\pi}{4}$. Then we generate the angles of arrival and
departure, whose realizations are given by
\begin{align}
\begin{aligned}
&\bm{\theta}_{\textrm{r}}\!=\!
[0.6833, 0.5937, 0.5982, 0.5309, 0.7593, 0.7719]^T\\
&\bm{\theta}_{\textrm{t}}\!=\!
[0.7468, 0.8778, 0.8219, 0.8823, 1.0332, 1.1444]^T.
\end{aligned}
\end{align}
\begin{table*}[ht]
\centering
\begin{tabular}{l*{6}{c}r}
$\mathrm{SNR}(\mathrm{dB})$       & -35 & -30 & -25 & -20 & -15 & -10 & -5 \\
\hline
Average mutual information           &7.248s&8.969s&8.728s& 8.775s&8.857s&8.918s&8.630s\\
Lower bound w. const. shift          &0.230s&0.159s&0.156s&0.197s&0.151s&0.156s&0.211s\\
Lower bound approx. w. const. shift  &0.028s&0.016s&0.014s&0.016s&0.012s&0.012s&0.020s\\
\end{tabular}
\vspace{-0.1cm}
\caption{Running times (in secs.) versus $\mathrm{SNR}$ for average mutual information and its approximations.}
\vspace{-0.5cm}
\end{table*}
For illustration purpose, the effective analog precoder $\bar{\bF}$ is obtained by Algorithm 1, and the effective digital precoder is set as $\bar{\bB}\!=\!\bI$.

The function values and running times for the average mutual information, the lower bound with a constant shift and the lower bound approximation with a constant shift are presented in Fig. 1 and Table I. The simulated curve is obtained by the Monte Carlo method, which computes the average mutual information using 3000 realizations of $\bH$ and $\bn$. From Fig. 1 and Table I, we have the following remarks:
\begin{enumerate}
  \item With a constant shift, the lower bound provides a very accurate approximation to the average mutual information in whole SNR regimes.
  \item The lower bound approximation plus a constant and the average mutual information match exactly at low and high SNR regimes, and their gap at medium SNR regime is less than 0.5bps/Hz in our case.
  \item The lower bound approximation consumes much lower computational time than the average mutual information and its lower bound, thus we design hybrid precoders by maximizing the lower bound approximation.
\end{enumerate}

\subsection{Example 2: Convergence of the Manifold-based Gradient Ascent Algorithm}
In this subsection, we consider a mmWave MIMO system with $N_\mathrm{r}\!=\!16$, $N_\mathrm{t}\!=\!64$, $N_\mathrm{rf}\!=\!4$ and $N_\mathrm{s}\!=\!4$. The number of physical propagation paths is set as $L=6$, and the SNR is given by $\mathrm{SNR}\!=\!-22.5 \mathrm{dB}$. The input signal is drawn from QPSK modulation. The mean angle of $\bm{\theta_{\textrm{r}}}$ is uniformly distributed over $[0,2\pi]$, i.e., $\bar{\theta}_\mathrm{r}\sim \mathrm{unif}(0,2\pi)$. In contrast, the mean angle of $\bm{\theta_{\textrm{t}}}$ is set as $\bar{\theta}_\mathrm{t}\!=\!\frac{\pi}{3}$.
The realizations of $\bm{\theta_{\textrm{r}}}$ and $\bm{\theta_{\textrm{t}}}$ are given by
\begin{align}
\begin{aligned}
&\bm{\theta}_{\textrm{r}}\!=\!
[4.6448, 4.7492, 4.9337, 4.8962, 5.3448, 4.4681]^T\\
&\bm{\theta}_{\textrm{t}}\!=\!
[0.8806, 1.4545, 0.8359, 1.1047, 1.2880, 0.8917]^T.
\end{aligned}
\end{align}
The initial point of the effective analog precoder $\bar{\bF}_{\!\mathrm{init}}$ is obtained by Algorithm 1, and the initial point of the effective digital precoder $\bar{\bB}_{\mathrm{init}}$ is set as the right singular vectors of $\bA_{\mathrm{t}}^H\bar{\bF}_{\!\mathrm{init}}$.

The evolution of the proposed manifold-based gradient ascent algorithm is shown in Fig. 2. For comparison, it also shows the
hybrid precoding with block coordinate ascent algorithm, and the average mutual information without hybrid precoding. The block coordinate ascent algorithm solves the hybrid precoding problem \eqref{L3} by optimizing $\pmb{\Phi}$ and $\bar{\bB}$ alternatively with initial point $(\bar{\bF}_{\!\mathrm{init}},\bar{\bB}_{\mathrm{init}})$. The effective analog and digital precoders in no hybrid precoding case are set as
\begin{align}
\bm{\bar{F}}\!=\!\sqrt{\frac{N_\mathrm{rf}}{N_\mathrm{t}}}
\begin{bmatrix}
    \bm{1} & \dots  & \bm{0} \\
    \vdots & \ddots & \vdots \\
    \bm{0} & \dots  & \bm{1}
\end{bmatrix},\;\;\bm{\bar{B}}\!=\!\bI.
\end{align}

From Fig. 2, we observe that our proposed manifold-based gradient ascent algorithm converges to 1.165 bps/Hz after 16 iterations
\begin{figure}[!htp]
  \begin{center}
  \includegraphics[scale=0.6]{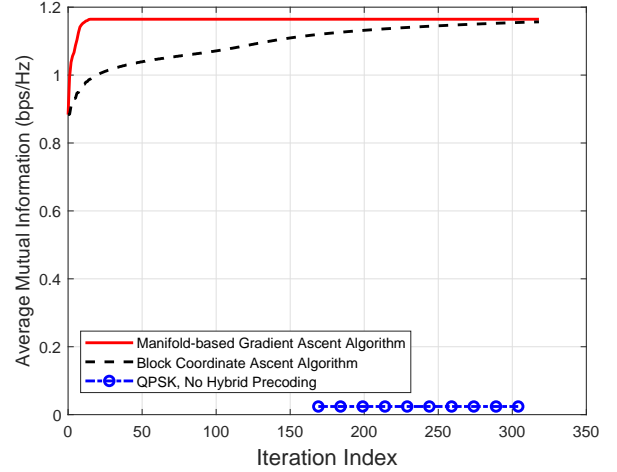}
  \vspace{-0.1cm}
  \caption{Evolution of average mutual information as the hybrid precoders are optimized with the proposed manifold-based gradient ascent and the block coordinate ascent algorithms. The input signal is drawn from QPSK; SNR is -22.5 dB.}
\end{center}
  \vspace{-0.8cm}
\end{figure}
while the block coordinate ascent algorithm requires over 320 iterations to approach the same value. Therefore, the proposed manifold-based gradient ascent algorithm is much faster than the block coordinate ascent algorithm. This phenomenon occurs mainly because our proposed algorithm updates $\pmb{\Phi}$ and $\bar{\bB}$ simultaneously while the block coordinate ascent algorithm updates $\pmb{\Phi}$ and $\bar{\bB}$ alternatively. In addition, we also observe that the performance of no hybrid precoding is very poor because we do not exploit any channel state information to design hybrid precoders.

The empirical cumulative distribution of average mutual information for the hybrid precoder from various initial points
of the effective analog and digital precoders are further depicted in Fig. 3, which is obtained by generating 3000 random initial points $(\bar{\bF}_{\!\mathrm{init}},\bar{\bB}_{\mathrm{init}})$. The initial analog precoders $\bar{\bF}_{\!\mathrm{init}}$ are obtained by Algorithm 1, whose output depends on the random input matrix $\bR$. The initial digital precoders $\bar{\bB}_{\mathrm{init}}$ are generated with i.i.d. zero-mean unit-variance complex Gaussian entries, and then normalized to satisfy the power constraint. The empirical cumulative distribution curve shows that although the hybrid precoding design with given partition of subsets is a nonconvex problem, our proposed manifold-based gradient ascent algorithm can achieve a near globally optimal solution from arbitrary initial points.

\subsection{Example 3: Performance of Mixed-CSI-based Hybrid Precoding}
In this subsection, we evaluate the performance of mixed-CSI-based hybrid precoding. We consider a mmWave MIMO system with $N_\mathrm{r}\!=\!24$, $N_\mathrm{t}\!=\!64$, $N_\mathrm{rf}\!=\!4$ and $N_\mathrm{s}\!=\!4$. The number of physical propagation paths is set as $L=8$. The input signal is drawn from BPSK modulation.
\begin{figure}[!htp]
  \begin{center}
  \includegraphics[scale=0.6]{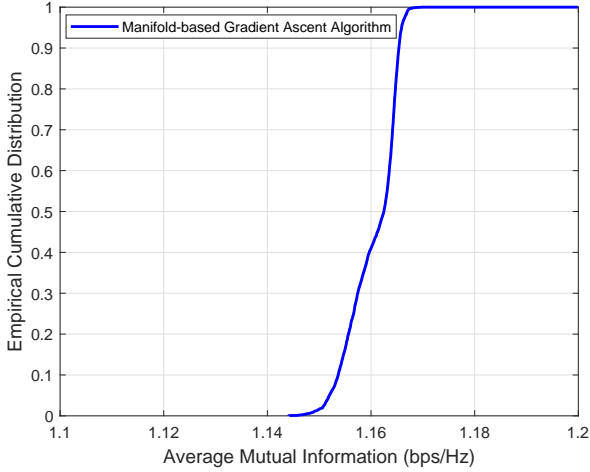}
  \vspace{-0.1cm}
  \caption{Empirical cumulative distribution of average mutual information for various initial points. The input signal is drawn from QPSK; SNR is -22.5 dB.}
  \end{center}
  \vspace{-0.8cm}
\end{figure}
The mean angle of $\bm{\theta_{\textrm{r}}}$ satisfies $\bar{\theta}_\mathrm{r}\sim \mathrm{unif}(0,2\pi)$, and the mean angle of $\bm{\theta_{\textrm{t}}}$ is set as $\bar{\theta}_\mathrm{t}\!=\!\frac{\pi}{4}$. The realizations of $\bm{\theta_{\textrm{r}}}$ and $\bm{\theta_{\textrm{t}}}$ are given by
\begin{align}
\begin{aligned}
&\bm{\theta}_{\textrm{r}}\!=\!
[3.921, 3.442, 3.550, 3.449, 3.514, 3.415, 3.314, 3.289]^T\\
&\bm{\theta}_{\textrm{t}}\!=\!
[0.760, 0.614, 0.674, 0.683, 0.916, 0.749, 0.831, 0.777]^T.
\end{aligned}
\end{align}

The mixed-CSI-based hybrid precoding utilizes channel statistics to design the effective analog precoder $\bar{\bF}$, and then design the effective digital precoder based on the instantaneous CSI. To evaluate the average mutual information, we generate $N\!=\!1500$ independent samples of the channel matrix
\begin{align}
\bH_i\!=\!\sqrt{\frac{N_\mathrm{r}N_\mathrm{t}}{L}}\bA_\textrm{r}\bm{\Gamma}_i\bA_\textrm{t}^{\!H},\;i=1,2,...,N.
\end{align}
Then we solve the digital precoding problem \eqref{J7} for each effective channel matrix $\bH_i\bar{\bF}$. Finally the average mutual information is given by
\begin{align}
\frac{1}{N}\sum_{i=1}^N C(\bH_i\bar{\bF})
\end{align}
where $C(\bH_i\bar{\bF})$ is the maximum mutual information for given channel matrix $\bH_i\bar{\bF}$.

We make comparisons between the mixed-CSI-based hybrid precoding under finite-alphabet (FA) inputs and three interesting scenarios, namely the optimal unconstrained precoder with FA inputs \cite{xiao2011globally}, the instantaneous-CSI-based hybrid precoding under FA inputs, and the instantaneous-CSI-based hybrid precoding under Gaussian inputs \cite{park2017dynamic}. All hybrid precoding algorithms are designed for the dynamic subarray structure. The instantaneous-CSI-based hybrid precoding with FA inputs first solve the following dynamic subarray problem
\begin{equation}\label{M1}
\begin{aligned}
\underset{\bar{\bF}\in \mathcal{F},\{\mathcal{S}_j\}\in \mathcal{S}}{\mathrm{maximize}}\quad \trace\left(\bar{\bF}^H\bH^H\bH\bar{\bF}\right).
\end{aligned}
\end{equation}
Note that problem \eqref{M1} has the same mathematical structure with problem \eqref{J6}, thus we can solve it using Algorithm 1.
\begin{figure}[!htp]
  \begin{center}
  \includegraphics[scale=0.6]{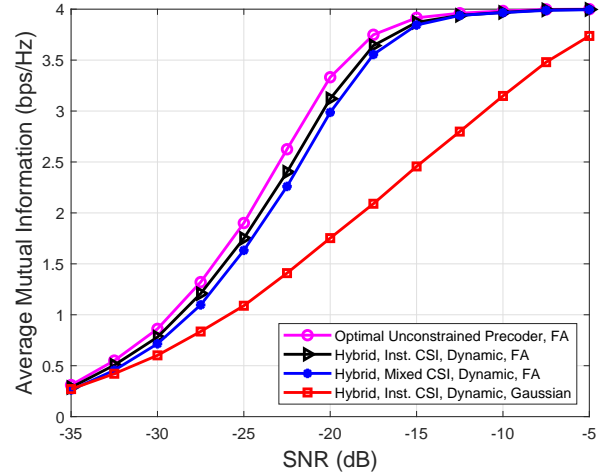}
  \vspace{-0.1cm}
  \caption{Average mutual information versus SNR for different scenarios in a mmWave MIMO channel ($N_\mathrm{r}\!=\!24$, $N_\mathrm{t}\!=\!64$, $N_\mathrm{rf}\!=\!4$, $N_\mathrm{s}\!=\!4$, $L=8$).}
\end{center}
  \vspace{-0.5cm}
\end{figure}
Then we design analog and digital precoders by maximizing the mutual information with given partition of subsets, i.e., we solve the following optimization problem
\begin{equation}\label{M2}
\begin{aligned}
\underset{\bar{\bF}\in \mathcal{F},\bar{\bB}\in \mathcal{B}}{\mathrm{maximize}}\quad \mathcal{I}(\bx;\by|\bH)
\end{aligned}
\end{equation}
using the manifold-based gradient ascent algorithm (Algorithm 2).

Fig. 4 demonstrates the average mutual information versus SNR for different scenarios. From Fig. 4, we have the following remarks:
\begin{enumerate}
  \item The performance of optimal unconstrained precoders is the benchmark for any hybrid precoding schemes, and the proposed hybrid precoding with dynamic subarrays has about  $1\mathrm{dB}$ performance loss compared with the optimal unconstrained precoder. Therefore, the the hybrid precoding with dynamic subarrays provides a good tradoff between performance and complexity.
  \item The performance gap between our proposed mixed-CSI-based hybrid precoding and the instantaneous-CSI-based hybrid precoding is very small, while the mixed-CSI-based hybrid precoding can greatly reduce the complexity of hybrid precoding design and channel estimation.
  \item The mixed-CSI-based hybrid precoding with finite-alphabet inputs can achieve 3.5bps/Hz when $\mathrm{SNR}=-17.5\mathrm{dB}$, while the instantaneous-CSI-based hybrid precoding under Gaussian inputs requires $-7.5\mathrm{dB}$ to approach the same value. Therefore, our proposed mixed-CSI-based hybrid precoding has a maximum $10\mathrm{dB}$ gain compared with the instantaneous-CSI-based hybrid precoding under Gaussian inputs. This is mainly because hybrid precoders designed under Gaussian inputs will lead to significant performance loss when applied to systems employing FA.
\end{enumerate}

\begin{figure}[h]
  \begin{center}
  \includegraphics[scale=0.6]{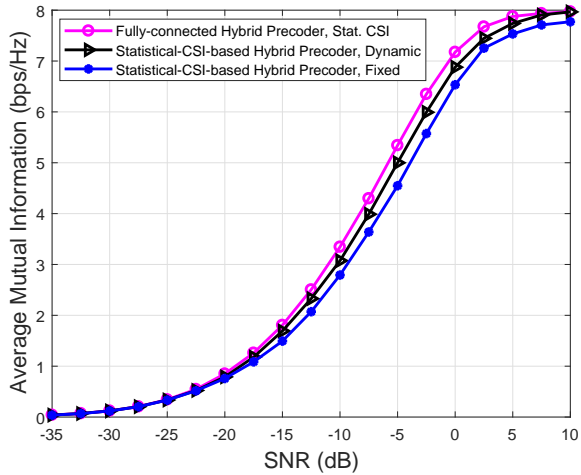}
  \vspace{-0.3cm}
  \caption{Average mutual information versus SNR for mmWave MIMO channel with $N_\mathrm{r}\!=\!4$, $N_\mathrm{t}\!=\!64$, $N_\mathrm{rf}\!=\!4$, $N_\mathrm{s}\!=\!4$, $L=5$.}
\end{center}
  \vspace{-0.7cm}
\end{figure}
\subsection{Example 4: Performance of Statistical-CSI-based Hybrid Precoding}
In this subsection, we  consider a mmWave MIMO system with $N_\mathrm{r}\!=\!4$, $N_\mathrm{t}\!=\!64$, $N_\mathrm{rf}\!=\!4$ and $N_\mathrm{s}\!=\!4$. The number of physical propagation paths is set as $L=5$. The input signal is drawn from QPSK modulation. The mean angle of $\bm{\theta_{\textrm{r}}}$ satisfies $\bar{\theta}_\mathrm{r}\sim \mathrm{unif}(0,2\pi)$, and the mean angle of $\bm{\theta_{\textrm{t}}}$ is set as $\bar{\theta}_\mathrm{t}\!=\!\frac{\pi}{4}$. The realizations of $\bm{\theta_{\textrm{r}}}$ and $\bm{\theta_{\textrm{t}}}$ are given by
\begin{align}
\begin{aligned}
&\bm{\theta}_{\textrm{r}}\!=\!
[0.4186, 0.5499, 0.4839, 0.3135, 0.7505]^T\\
&\bm{\theta}_{\textrm{t}}\!=\!
[0.9144, 0.7117, 0.7969, 0.8150, 0.6860]^T.
\end{aligned}
\end{align}

We first evaluate the spectral efficiency of the statistical-CSI-based hybrid precoding with dynamic subarrays. We set the fully-connected hybrid precoding under statistical CSI as the benchmark, and then make comparisons between the statistical-CSI-based hybrid precoding with dynamic subarrays and statistical-CSI-based hybrid precoding with fixed subarrays. All hybrid precoding algorithms are designed for FA inputs. The fully-connected hybrid precoder under statistical CSI factorizes the optimal unconstrained precoder into analog and digital precoders \cite{jin2018hybrid}, and the optimal unconstrained precoder can be obtained by maximizing the lower bound approximation with projected gradient algorithm \cite{boyd2004convex}. The statistical-CSI-based hybrid precoding with fixed subarrays utilizes Algorithm 2 to solve problem \eqref{L3} with the following given $\{\mathcal{S}_j\}$:
\begin{align}
\mathcal{S}_j\!=\!\big\{(j-1)q\!+\!1,(j-1)q\!+\!2,...,(j-1)q\!+\!q\big\}, \;\;\forall j
\end{align}
\begin{figure}[h]
  \begin{center}
  \includegraphics[scale=0.6]{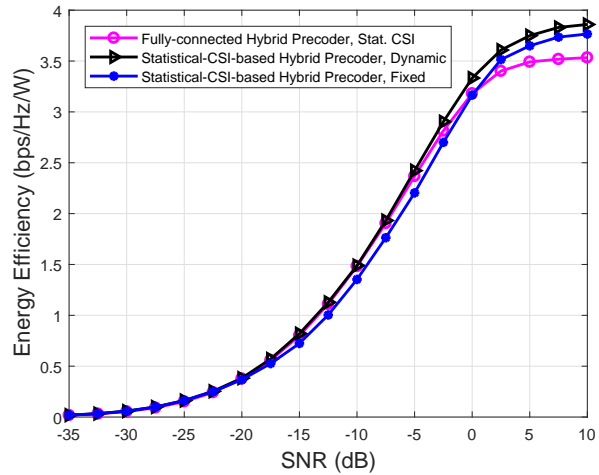}
  \vspace{-0.3cm}
  \caption{Energy efficiency versus SNR for mmWave MIMO channel with $N_\mathrm{r}\!=\!4$, $N_\mathrm{t}\!=\!64$, $N_\mathrm{rf}\!=\!4$, $N_\mathrm{s}\!=\!4$, $L=5$.}
\end{center}
  \vspace{-0.8cm}
\end{figure}
where $q\!=\!\frac{N_\mathrm{t}}{N_\mathrm{rf}}$. The results in Fig. 5 show that the statistical-CSI-based hybrid precoding with dynamic subarrays has about 1dB performance gain over the statistical-CSI-based hybrid precoding with fixed subarrays in the medium and high SNR regimes.

Then we evaluate the energy efficiency of the statistical-CSI-based hybrid precoding with dynamic subarrays. Based on the energy consumption model in \cite{cui2005energy}, the energy efficiency $\eta$ is defined as
\begin{align}
\eta=\frac{E_{\bH}\mathcal{I}(\bx;\by|\bH)}{P+N_{\mathrm{rf}}P_{\!\mathrm{rf}}+N_{\mathrm{ps}}P_{\!\mathrm{ps}}}
\end{align}
where $P$ is the transmit power, $P_{\!\mathrm{rf}}$ is the power consumed by RF chain, $P_{\!\mathrm{ps}}$ is the power consumed by phase shifter, $N_{\mathrm{rf}}$ and $N_{\mathrm{ps}}$ are the
numbers of required RF chains and phase shifters, respectively. In this paper, we use the practical values $P_{\!\mathrm{rf}}=250 \mathrm{mW}$ \cite{amadori2015low},
$P_{\!\mathrm{ps}}=1 \mathrm{mW}$\cite{constantine2012antenna}, and $P=1 \mathrm{W}$ (about 30 dBm) in a small cell transmission scenario \cite{rappaport2011state}. Fig. 6 shows the energy efficiency
comparison for the fully-connected hybrid precoding under statistical CSI as well as the statistical-CSI-based hybrid precoding with dynamic and fixed subarrays. We observe that our proposed statistical-CSI-based hybrid precoding with dynamic subarrays outperforms the fully-connected hybrid precoding under statistical CSI in the high SNR regime.

\section{Conclusion}
In this paper, we have considered the partially-connected hybrid precoding design for millimeter wave (mmWave) multiple-input multiple-output (MIMO) systems with  finite-alphabet inputs and dynamic subarrays. The analog and digital precoders are designed using either statistical CSI or mixed CSI. To simplify the original problem, we have proposed a  low complexity algorithm to design the near-optimal partition of antennas using statistical CSI. Then a lower bound and its approximation have been derived for the
average mutual information. The lower bound plus a constant offers a very accurate approximation to the average mutual information, and the computational complexity of the lower bound and its approximation are a few orders of magnitudes less than that of the average mutual information. Furthermore, a manifold-based gradient ascent algorithm has been proposed to find optimal analog and digital precoders via maximizing the lower bound approximation of the average mutual information. Several numerical results have also been provided to show that our proposed algorithm outperforms existing hybrid precoding algorithms.

\section*{Appendix}
\begin{IEEEproof}[Proof of Proposition 1]
We start with the necessary condition, i.e., if $\bF$ is a feasible point of problem \eqref{K1}, then $\bF$ satisfies \eqref{K2}. Since $|\bF_{\!ij}|\!=\!\mathbf{1}_{\mathcal{S}_j}(i)$, we have $|\bF_{\!ij}|\!\in\!\{0,1\}$. In addition, the following equations
\begin{align}
&\bigcup_{j=1}^{N_\mathrm{rf}}\mathcal{S}_j\!=\!\left\{1,2,...,N_\mathrm{t}\right\}\\
&\mathcal{S}_j \cap \mathcal{S}_k\!=\!\emptyset,\;\;\forall j\!\neq\! k
\end{align}
implies $\sum_{i=1}^{N_\mathrm{t}}\|\bF_{\!i\bullet}\|_0\!=\!N_\mathrm{t}$ and $\|\bF_{\!i\bullet}\|_0\leq 1$, respectively. Therefore, we have $\|\bF_{\!i\bullet}\|_0\!=\!1$. This completes the first part of the proof.

Next, we prove the sufficient condition, i.e., if $\bF$ satisfies \eqref{K2}, then $\bF$ is a feasible point of problem \eqref{K1}. Let $\mathcal{S}_j$ denotes positions of nonzero entries in the $j$th column of $\bF$. Since $|\bF_{\!ij}|\!\in\!\{0,1\}$, we can express $\bF$ as
\begin{align}
|\bF_{\!ij}|\!=\!\left\{
\begin{aligned}
&1\quad\quad \mathrm{if}\;\;i\!\in\!\mathcal{S}_j \\
&0\quad\quad \mathrm{otherwise}.
\end{aligned}
\right.
\end{align}
In addition, since $||\bF_{\!i\bullet}||_0\!=\!1$ for $i\!\in\!\{1,2,...,n_\mathrm{T}\}$, $\{\mathcal{S}_j\}$ must satisfy
\begin{align}
&\bigcup_{j=1}^{N_\mathrm{rf}}\mathcal{S}_j\!=\!\big\{1,2,...,N_\mathrm{t}\big\}\\
&\mathcal{S}_j \cap \mathcal{S}_\ell\!=\!\emptyset,\;\;\forall j\!\neq\! \ell.
\end{align}
Therefore, $\{\mathcal{S}_j\}_{j=1}^{n_\mathrm{RF}}\in \mathcal{S}$ and this completes the proof.
\end{IEEEproof}

\begin{IEEEproof}[Proof of Proposition 2]
Note that $\log(x)$ is a concave function for $x\!>\!0$. Using Jensen's inequality, the average mutual information with finite-alphabet inputs $E_{\bH}\mathcal{I}(\bx;\by|\bH)$ can be lower bounded by
\begin{align}\label{A1}
E_{\bH}\mathcal{I}(\bx;\by|\bH)\!\geq\!\log& K\!-\!N_\textrm{r}\left(\frac{1}{\ln 2}\!-\!1\right)\!-\!\frac{1}{K}\sum_{m=1}^{K}\nonumber\\
&\log\!\sum_{k=1}^{K}E_{\bH,\bn}\!\exp\left(\!-\frac{\|\be_{mk}\!+\!\bn\|^2}{\sigma^2}\right)
\end{align}
where $\be_{mk}\!=\!\bH\bar{\bF}\bar{\bB}(\bx_m\!-\!\bx_k)$.

Since $\bn$ is the i.i.d. complex Gaussian noise, the expectation over $\bn$ in \eqref{A1} can be calculated as
\begin{align}\label{A2}
E_{\bn}&\!\exp\left(\!-\frac{\|\be_{mk}\!+\!\bn\|^2}{\sigma^2}\right)\nonumber\\
&\!=\!\frac{1}{(\pi\sigma^2)^{N_\textrm{r}}}\int_{\bn}\exp\left(\!-\frac{\|\be_{mk}\!+\!\bn\|^2\!+\!\|\bn\|^2}{\sigma^2}\right)\mathrm{d}\bn\nonumber\\
&\!=\!\frac{1}{(\pi\sigma^2)^{N_\textrm{r}}}\int_{\bn}\prod_{i=1}^{N_\textrm{r}}\exp\left(\!-\frac{|e_{mk,i}\!+\!n_i|^2\!+\!|n_i|^2}{\sigma^2}\right)\mathrm{d}\bn\nonumber\\
&\!=\!\prod_{i=1}^{N_\textrm{r}}\frac{1}{\pi\sigma^2}\int_{n_i}\exp\left(\!-\frac{|e_{mk,i}\!+\!n_i|^2\!+\!|n_i|^2}{\sigma^2}\right)\mathrm{d}n_i
\end{align}
where $e_{mk,i}$ and $n_i$ are the $i$th element of $\be_{mk}$ and $\bn$, respectively. Applying the integrals of exponential function and extending it to the complex-valued case, equation \eqref{A2} is rewritten as
\begin{align}\label{A3}
E_{\bn}\!\exp\left(\!-\frac{\|\be_{mk}\!+\!\bn\|^2}{\sigma^2}\right)&\!=\!\prod_{i=1}^{N_\textrm{r}}\frac{1}{2}\exp\left(\!-\frac{|e_{mk,i}|^2}{2\sigma^2}\right)\nonumber\\
&\!=\!\frac{1}{2^{N_\textrm{r}}}\exp\left(\!-\frac{\be_{mk}^H\be_{mk}}{2\sigma^2}\right).
\end{align}
Then we insert $\be_{mk}\!=\!\left(\frac{N_\mathrm{r}N_\mathrm{t}}{L}\right)^{\frac{1}{2}}\bA_\textrm{r}\bm{\Gamma}\bA_\textrm{t}^{\!H}\bar{\bF}\bar{\bB}(\bx_m\!-\!\bx_k)$ into equation \eqref{A3}, and it yields
\begin{align}\label{A4}
\exp&\left(\!-\frac{\be_{mk}^H\be_{mk}}{2\sigma^2}\right)\!=\!\exp\left[-\trace\left(\bm{\Gamma}\bW_{\!mk}\bm{\Gamma}^H\bA_\mathrm{r}^H\bA_\mathrm{r}\right)\right]
\end{align}
where $\bW_{\!mk}\!=\!\frac{N_\mathrm{r}N_\mathrm{t}}{2\sigma^2L}\bA_\textrm{t}^{\!H}\bar{\bF}\bar{\bB}(\bx_m\!-\!\bx_k)(\bx_m\!-\!\bx_k)^H\bar{\bB}^H\bar{\bF}^H\bA_\textrm{t}$. Since $\bm{\Gamma}$ is a diagonal matrix, we have
\begin{align}
\trace\left(\bm{\Gamma}\bW_{\!mk}\bm{\Gamma}^H\bA_\textrm{r}^H\bA_\textrm{r}\right)\!=\!\bd^H\left[\left(\bA_\textrm{r}^H\bA_\textrm{r}\right)^T\!\circ\!\bW_{\!mk}\right]\bd
\end{align}
where $\bd$ is the diagonal entries of $\bm{\Gamma}^*$. Since the diagonal entries of $\bm{\Gamma}$ are i.i.d. complex Gaussian distributed, $\bd$ is an i.i.d. complex Gaussian vector. Then the expectation of equation \eqref{A4} over $\bm{\Gamma}$ can be expressed as
\begin{align}\label{A5}
E_{\bm{\Gamma}}&\exp\left[-\trace\left(\bm{\Gamma}\bW_{\!mk}\bm{\Gamma}^H\bA_\textrm{r}^H\bA_\textrm{r}\right)\right]\nonumber\\
&\!=\!E_{\bd}\exp\left[-\bd^H\left(\bA_\textrm{r}^H\bA_\textrm{r}\right)^T\!\circ\!\bW_{\!mk}\bd\right]\nonumber\\
&\!=\!\frac{1}{\pi^L}\int_{\bd}\exp\left[-\bd^H\left(\bI\!+\!\left(\bA_\textrm{r}^H\bA_\textrm{r}\right)^T\!\circ\!\bW_{\!mk}\right)\bd\right]\mathrm{d}\bd\nonumber\\
&\!=\!\det\left[\bI\!+\!\left(\bA_\textrm{r}^H\bA_\textrm{r}\right)^T\!\circ\!\bW_{\!mk}\right]^{-1}.
\end{align}
The combination of \eqref{A1}, \eqref{A3}, \eqref{A4} and \eqref{A5} yields the lower bound. This completes the proof.
\end{IEEEproof}

\begin{IEEEproof}[Proof of Proposition 3]
Since $\lim_{N_\mathrm{r}\rightarrow \infty}\bA_\textrm{r}^H\bA_\textrm{r}\!=\!\bI$, we can replace $\bA_\textrm{r}^H\bA_\textrm{r}$ by $\bI$. Then $L(\bar{\bF},\bar{\bB})$ can be approximated as
\begin{align}\label{A6}
L(\bar{\bF},\bar{\bB})\!\approx\!\log& K\!-\!N_\textrm{r}\left(\frac{1}{\ln 2}\!-\!1\right)\nonumber\\
&\!-\!\frac{1}{K}\sum_{m=1}^{K}\log\sum_{k=1}^K\det\Big[\bI\!+\!\bI\!\circ\!\bW_{\!mk}\Big]^{-1}.
\end{align}
Note that $\bI+\bI\circ\bW_{\!mk}$ is a diagonal matrix, with the $\ell$th diagonal element being
\begin{align}\label{A7}
\Big[\bI\!+\!\bI\!\circ\!\bW_{\!mk}\Big]_{\ell\ell}\!=\!1+\frac{N_\mathrm{r}N_\mathrm{t}}{2\sigma^2L}\left|\beta_{mk\ell}\right|^2
\end{align}
where $\beta_{mk\ell}\!=\!\ba(\theta_{\textrm{t},\ell})^H\bar{\bF}\bar{\bB}(\bx_m\!-\!\bx_k)$. Combining \eqref{A6} and \eqref{A7} yields the lower bound approximation. This completes the proof.
\end{IEEEproof}

\begin{IEEEproof}[Proof of Proposition 4]
We first rewrite $L_{\!A}(\bar{\bF},\bar{\bB})$ as
\begin{align}
L_{\!A}(\bar{\bF},\bar{\bB})\!=\!\log& K\!-\!N_\textrm{r}\left(\frac{1}{\ln 2}\!-\!1\right)\!-\!\frac{1}{K}\sum_{m=1}^{K}\log\nonumber\\
&\sum_{k=1}^K\exp\left[-\!\sum_{\ell=1}^{L}\ln\left(1\!+\!\frac{N_\mathrm{r}N_\mathrm{t}}{2\sigma^2L}\left|\beta_{mk\ell}\right|^2\right)\right].
\end{align}
Using the chain rule in differentiation, the differential of $\bar{\mathcal{I}}_A(\bar{\bF},\bar{\bB})$ with respect to $\bar{\bP}\!=\!\bar{\bF}\bar{\bB}$ is
\begin{align}\label{L4}
\mathrm{d}L_{\!A}\!=\!\trace&\left(\mathrm{d}\bar{\bP}^H\!\bG\!+\!\bG^H\mathrm{d}\bar{\bP}\right)
\end{align}
where $\bG\!=\!\sum_{\ell=1}^L \ba(\theta_{\textrm{t},\ell})\ba(\theta_{\textrm{t},\ell})^H\bar{\bF}\bar{\bB}\bE_{\ell}$. Inserting $\mathrm{d}\bar{\bP}\!=\!\bar{\bF}\mathrm{d}\bar{\bB}$ into equation \eqref{L4}, we obtain
\begin{align}\label{L5}
\mathrm{d}L_{\!A}\!=\!\left(\mathrm{d}\bar{\bB}^H\bar{\bF}^H\bG
\!+\!\bG^H\bar{\bF}\mathrm{d}\bar{\bB}\right).
\end{align}
Since $R(\pmb{\Phi},\bar{\bB})\!\triangleq\!L_{\!A}(\bar{\bF}(\pmb{\Phi}),\bar{\bB})$, the gradient of $R(\pmb{\Phi},\bar{\bB})$ with respect to $\bar{\bB}$ is
\begin{align}
\nabla_{\!\bar{\bB}}R(\pmb{\Phi},\bar{\bB})\!=\!\bar{\bF}^H\bG.
\end{align}

Similarly, inserting $\mathrm{d}\bar{\bP}\!=\!\mathrm{d}\bar{\bF}\bar{\bB}$ into equation \eqref{L4}, we obtain
\begin{align}\label{L6}
\mathrm{d}L_{\!A}\!=\!\trace&\left(\mathrm{d}\bar{\bF}^H\!\bG\bar{\bB}^H\!+\!\bar{\bB}\bG^H\mathrm{d}\bar{\bF}\right).
\end{align}
In addition, since $\mathrm{d}\bar{\bF}_{\!ij}\!=\!\jmath\bar{\bF}_{\!ij}\mathrm{d}\pmb{\Phi}_{\!ij}$, the differential of $\bar{\bF}$ is given by
\begin{align}
\mathrm{d}\bar{\bF}\!=\!\jmath\mathrm{d}\pmb{\Phi}\circ\bar{\bF}.
\end{align}
Inserting $\mathrm{d}\bar{\bF}\!=\!\jmath\mathrm{d}\pmb{\Phi}\circ\bar{\bF}$ into equation \eqref{L6} and using the following two equations
\begin{align}
&R(\pmb{\Phi},\bar{\bB})\!\triangleq\!L_{\!A}(\bar{\bF}(\pmb{\Phi}),\bar{\bB})\\
&\trace\left[(\bA\circ \bB)\bC\right]\!=\!\trace\left[\bA(\bC\circ \bB^T)\right]
\end{align}
we conclude that
\begin{align}
\nabla_{\!\pmb{\Phi}}R(\pmb{\Phi},\bar{\bB})\!=\!2\Im\left(\bG\bar{\bB}^H\!\circ\! \bar{\bF}^*\right).
\end{align}
This completes the proof.
\end{IEEEproof}

\maketitle
\bibliographystyle{IEEEtran}
\bibliography{IEEEabrv,reference}

\end{document}